\documentclass[final,3p,times]{elsarticle}

\usepackage{amssymb}

\usepackage{amssymb}
\usepackage{amsmath}
\usepackage{mathtools}
\usepackage[normalem]{ulem}
\usepackage{subfig}

\newcommand{\flr}[1]{\lfloor #1 \rfloor}
\newtheorem{problem}{Problem}
\DeclarePairedDelimiter{\abs}{\lvert}{\rvert}

\journal{Nuclear Physics B}

\usepackage{color}
\definecolor{forestgreen}{rgb}{0.33,0.61,0.34}

\mathtoolsset{showonlyrefs=true}
\begin{document}

\begin{frontmatter}



\title{Cyclic Pursuit Formation Control for Arbitrary Desired Shapes}


\author[OU]{Anna Fujioka}

\author[HU,OU]{Masaki Ogura}

\affiliation[OU]{organization={Osaka University},
            addressline={1-5 Yamadaoka}, 
            city={Suita},
            postcode={565-0871}, 
            state={Osaka},
            country={Japan}}

\affiliation[HU]{organization={Hiroshima University},
            addressline={1-7-1 Kagamiyama}, 
            city={Higashi Hiroshima},
            postcode={739-8521}, 
            state={Hiroshima},
            country={Japan}}

\author[OU]{Naoki Wakamiya}

\begin{abstract}
A multi-agent system comprises numerous agents that autonomously make decisions to collectively accomplish tasks, drawing significant attention for their wide-ranging applications. Within this context, formation control emerges as a prominent task, wherein agents collaboratively shape and maneuver while preserving formation integrity. Our focus centers on cyclic pursuit, a method facilitating the formation of circles, ellipses, and figure-eights under the assumption that agents can only perceive the relative positions of those preceding them. However, this method's scope has been restricted to these specific shapes, leaving the feasibility of forming other shapes uncertain. In response, our study proposes a novel method based on cyclic pursuit capable of forming a broader array of shapes, enabling agents to individually shape while pursuing preceding agents, thereby extending the repertoire of achievable formations. We present two scenarios concerning the information available to agents and devise formation control methods tailored to each scenario. Through extensive simulations, we demonstrate the efficacy of our proposed method in forming multiple shapes, including those represented as Fourier series, thereby underscoring the versatility and effectiveness of our approach.
\end{abstract}



\begin{keyword}
Cyclic pursuit \sep formation control \sep Fréchet distance 



\end{keyword}

\end{frontmatter}


\section{Introduction}\label{sc:intro}

A multi-agent system (MAS) constitutes a sophisticated framework wherein multiple autonomous agents collaborate to make collective decisions and execute designated tasks, as extensively studied in the literature~\cite{olfati2007consensus}. Its inherent ability to tackle challenges surpassing the capacity of individual agents, coupled with its robustness against failures and disturbances, renders it applicable across diverse domains, ranging from robotics to social networks~\cite{siljak2022cyborg, kegeleirs2021swarm}. Among the myriad of tasks within MAS, formation control emerges as a quintessential endeavor, entailing multiple agents orchestrating themselves into various formations while preserving the integrity of the ensemble, a task that has garnered considerable research interest in recent years~\cite{oh2015survey}.

Despite the plethora of control algorithms proposed for multi-agent tasks, many of these approaches hinge on the premise of dense interaction dynamics, necessitating rigid interaction topologies among agents to achieve desired outcomes effectively~\cite{de2016distributed, hauri2013multi}. For instance, the method proposed by De Marina et al. underscores the importance of a rigid interaction topology for successful task execution~\cite{de2016distributed}.

In contrast, the cyclic pursuit method stands out for its remarkable ability to achieve formation control with limited information, relying solely on the relative position of the agent ahead, thus offering a more flexible and scalable approach to formation control within MAS~\cite{bruckstein1991ant, bernhart1959polygons, watton1969analytical, bruckstein1991ants2}. Originating as an attempt to mimic biological entities such as dogs and ants, cyclic pursuit has since evolved into a versatile approach known as the ``bugs'' problem, which has garnered considerable attention in both academic and industrial circles~\cite{bruckstein1991ant, bernhart1959polygons}. Previous studies have extensively explored its dynamics across various agent models, encompassing ants, crickets, frogs, and others, further solidifying its status as a cornerstone of formation control within MAS~\cite{bruckstein1991ants2, marshall2003pursuit}.

Motivated by the question of whether cyclic pursuit enables agents to form shapes beyond those previously demonstrated, this study proposes a novel method for forming desired shapes based on the cyclic pursuit strategy. Addressing two distinct problem settings, we investigate the sufficiency of information for agents to achieve formation for any given shape. In the first setting, agents share a coordinate system, thereby enhancing coordination and communication among agents. In contrast, in the second setting, agents do not share a coordinate system, thus imposing additional challenges that need to be addressed for successful formation control.

Our proposed method leverages cyclic pursuit dynamics, offering a systematic and scalable approach to formation control within MAS. In the first setting, agents calculate their ideal positions based on the preceding agent's position and move accordingly, iteratively forming the required shape. In the second setting, an additional operation is introduced to align the angles of coordinate systems, enabling agents to synchronize their angles with the movement direction of the predecessor agent, thus facilitating smoother coordination among agents.

This paper contributes to the advancement of formation control theory in MAS by showcasing the capability of the cyclic pursuit strategy to realize various shapes, surpassing its previously perceived limitations. Additionally, our work sheds light on the intricate dynamics of cyclic pursuit, paving the way for future research in formation control using cost-effective agents and expanding the boundaries of what is achievable within MAS.

The structure of this paper is as follows. Section~\ref{sc:statement} formulates the problem studied herein, elucidating the significance and scope of the research endeavor. In Section~\ref{sc:method}, we delve into the intricacies of the proposed methods, offering detailed insights into the underlying principles and algorithms employed. Subsequently, in Section~\ref{sc:sim}, we meticulously evaluate the effectiveness of the proposed methods through extensive simulations, providing empirical evidence to support our claims and hypotheses. Finally, Section~\ref{sc:conclusion} provides a comprehensive summary of our findings, outlines avenues for future research, and offers concluding remarks on the implications of our work for the broader field of multi-agent systems and formation control.
\section{Problem Statement}\label{sc:statement}

In this section, we describe the problem studied in this paper. We consider a multi-agent system in a two-dimensional plane~${\mathbb R}^2$. We assume that the system consists of $N$ agents. We assign the numbers $1$, \dots, $N$ to the agents. In addition, we let $x_i(k)$ denote the position of the $i$th agent at time~$k=0,1,2,\dotsc$.
Additionally, we define the local coordinate system that each agent possesses. Specifically, the rotation of the local coordinate system of agent $i$ at step $k$ relative to the global coordinate system (i.e. the $xy$--axis) is denoted as $\theta_i(k)$.
Under these assumptions, the objective of the multi-agent system is to form a complete required shape by the agents' distributed decision. In this paper, the desired shape is assumed to be described by a closed curve~$\gamma\colon [0, 1]\to \mathbb R^2$. For example, if the desired shape is a square, then we can use $\gamma$ given by 
\begin{equation}\label{eq:gamma_square}
\gamma(t) =
\begin{cases}
\displaystyle 
(1, 4t)
, &\mbox{if $t < \frac{1}{4}$}, 
\\
(1-4(t-\frac{1}{4}),1)
, &\mbox{if $\frac{1}{4}\leq t < \frac{1}{2}$},
\\
(0, 1-4(t-\frac{1}{2}))
, &\mbox{if $\frac{1}{2}\leq t < \frac{3}{4}$},
\\
(4(t-\frac{3}{4}), 0)
, &\mbox{otherwise}.
\end{cases}
\end{equation}

It is well-established~\cite{marshall2006pursuit} that the aforementioned objective can be achieved by a multi-agent system performing a cyclic pursuit when $\gamma$ is the unit circle, that is, the system converges to the state where all the agents form a circle by moving on it with the same speed while maintaining equal distances between two consecutive agents. Notably, the circle formed by the agents does not necessarily coincide with the unit circle specified by~$\gamma$. Therefore, the cyclic pursuit strategy guarantees the existence of a closed curve~$\tilde \gamma$ representing a circle, potentially different from~$\gamma$, and functions~$\tau_1$, \dots, $\tau_N$ from the set~$\{0, 1, 2, \dotsc\}$ of nonnegative integers to the interval~$[0,1]$ such that 
\begin{align}\label{eq:position_lim}
    \lim_{k \to \infty} \bigl(x_i(k) - \tilde{\gamma}(\tau_i(k))\bigr) = 0
\end{align}
and 
\begin{align}\label{eq:tau_lim}
   \lim_{k\to\infty}\delta\bigl(\tau_{n(i)}(k), \tau_i(k)\bigr) = \frac{1}{N}
\end{align}
for all $i=1, \dotsc, N$, where $n(i)$ represents the index of the agent~$i$'s predecessor agent and is defined by
\begin{equation}\label{eq:next_i}
n(i) =
\begin{cases}
\displaystyle 
i+1
, &\mbox{if $i < N$}, 
\\
1
, &\mbox{otherwise},
\end{cases}
\end{equation} 
and $\delta$ denotes the distance function equipped on the space $[0, 1)$ defined by 
\begin{equation}\label{eq:def_distance}
    \delta(x, y) = \min\bigl(\abs{x-y}, \abs{x+1-y}\bigr)
\end{equation}
for all $x, y\in [0, 1)$.

The first equation~\eqref{eq:position_lim} implies that, asymptotically, an agent~$i$ will move on $\tilde \gamma$ with its position specified by~$\tau_i(k)$. The second equation~\eqref{eq:tau_lim} implies that, under the parametrization of~$\tilde \gamma$ by~$[0, 1]$, any adjacent pair $(i, n(i))$ of agents travels with an equal distance of $1/N$ (i.e., the length of interval $[0, 1]$ divided by the number of agents) after a sufficient time from the start of the formation control.

This observation leads to the following question: can a multi-agent system with cyclic pursuit-based interactions among agents realize a formation other than a specific shape? 
Furthermore, what level of information suffices to achieve such formations? To delve into these questions, we delineate two distinct problem settings. In our first setting, we establish the agents' dynamics on cyclic pursuit principles. Consequently, each agent must ascertain its individual movement vector based on information pertaining to the agent directly ahead. Specifically, we presume that
\begin{description}
\item[A1)] each agent is able to observe the relative position of its predecessor agent.
\end{description}
In addition to the information A1), we assume that 
\begin{description}
\item[A2)] each agent has knowledge of the desired shape~$\gamma$;
\item[A3)] each agent~$i$ is aware of its own orientation $\phi_i(k)\in[0, 2\pi)$ (i.e., the angle of the direction vector measured as a counter-clockwise rotation from the positive $x$ axis) at any step $k$.
\end{description}
Within the framework of these assumptions, we formulate the first problem under investigation in this study as follows:
\begin{problem}
Let $\gamma$ be a closed curve. Under assumptions A1-A3, Design a distributed movement law for each agent such that the positions~$x_i$ of all the agents satisfy Eq.~\eqref{eq:position_lim} for a closed curve~$\tilde \gamma$ similar to $\gamma$ (i.e. enlarged and translated version of the desired shape $\gamma$) and a set of functions $\tau_i\colon \{0, 1, 2, \dotsc \} \to [0, 1)$ ($i=1, \dotsc, N$) that satisfy Eq.~\eqref{eq:tau_lim}. 
\end{problem}

In the second problem setting, along with the information A1) and A2), we assume that 
\begin{description}
\item[A4)] each agent has access to the 1-step previous positions of its predecessor agent;
\item[A5)] each agent~$i$ is aware of its own orientation $\phi_i(0)\in[0, 2\pi)$ (i.e., the angle of the direction vector measured as a counter-clockwise rotation from the positive $x$ axis) at the first step $k=0$.
\end{description}
Similar to problem 1, we formulate the second problem as follows:
\begin{problem}
Let $\gamma$ be a closed curve. Under assumptions A1, A2, A4, and A5, design a distributed movement law for each agent such that the positions~$x_i$ of all the agents satisfy Eq.~\eqref{eq:position_lim} for a closed curve~$\tilde \gamma$ (enlarged, translated and rotated version of the desired shape $\gamma$) and a set of functions $\tau_i\colon \{0, 1, 2, \dotsc \} \to [0, 1)$ ($i=1, \dotsc, N$) that satisfy Eq.~\eqref{eq:tau_lim}.
\end{problem}
\section{Proposed Method}\label{sc:method}

In this section, we present our solutions to Problems~1 and~2 that we posed in Section~\ref{sc:statement}. In the proposed methods, we assume that agents are evenly spaced on a circle with center~$[a, b]^\top$ and radius~$r$ at the initial step $k=0$. This assumption is feasible using conventional cyclic pursuit methods. In other words, it can be achieved if agents can obtain the relative positions of leading agents. We remark that both problem scenarios addressed in this research set the relative position acquisition as feasible under Assumption~A1, making the assumption achievable in either case.
Under this assumption, we can geometrically see that the difference in orientation between the consecutive agents equals $2\pi/N$. Therefore, the following equation holds for any agent~$i$:
\begin{equation}\label{eq:phi_equal}
\phi_{n(i)}(0) - \phi_i(0) =
\begin{cases}
\displaystyle 
\frac{2\pi}{N}
, &\mbox{if $\phi_i(0)<\phi_{n(i)}(0)$}, 
\vspace{3mm}
\\
\displaystyle 
\frac{2\pi}{N}-2\pi
, &\mbox{otherwise}.
\end{cases}
\end{equation}

We then describe the configuration of $\tau_i$ such that the Eq.~\eqref{eq:tau_lim} is satisfied. Under assumption~A3 for the first problem setting and assumption~A5 for the second problem setting, we let agent $i$ initialize its own internal variable~$\tau_i$ as
\begin{equation}\label{eq:tau_first}
    \tau_i(0) = \frac{\phi_i(0)}{2\pi}.
\end{equation}
Subsequently, we let each agent $i$ update $\tau_i$ at each time~$k$ as
\begin{equation}\label{eq:tau_update}
\tau_i(k+1) =
\begin{cases}
\displaystyle 
\tau_i(k) + \eta
, &\mbox{if $\tau_i(k) + \eta < 1$}, 
\\
\displaystyle 
\tau_i(k) + \eta - 1
, &\mbox{otherwise},
\end{cases}
\end{equation}
where $\eta\in(0,1)$ is a tunable parameter. From Eq.~\eqref{eq:phi_equal}, we can easily confirm that $\tau_i$ satisfies 
\begin{equation}\label{eq:distance1/N}
d(\tau_{n(i)}(k), \tau_i(k)) = 1/N    
\end{equation}
for all $k$. Therefore, Eq.~\eqref{eq:tau_lim} is satisfied. 

\subsection{Method 1}\label{subsc:method1}

In this section, we propose a movement law of the agents for Problem~1. As mentioned in Section~\ref{sc:method}, the objective in Problem~1 is to design a movement law of agents such that Eqs.~\eqref{eq:position_lim} and \eqref{eq:tau_lim} are satisfied under the assumptions~A1--A3. Since Eq.~\eqref{eq:tau_lim} is already satisfied as mentioned above, let us now focus on Eq.~\eqref{eq:position_lim}.

We first note that one of the potential movement laws of an agent, leading to the realization of Eq.\eqref{eq:position_lim}, can be expressed as
\begin{equation}\label{eq:idealmovement}
x_i(k) = \gamma(\tau_i(k))    
\end{equation}
However, this movement law cannot be practically implemented because it requires agents to have access to their own absolute positions. Since Problem~1 does not assume the availability of absolute coordinates, agents are unable to move by following \eqref{eq:idealmovement}. Consequently, within the proposed algorithm, we allow each agent to determine its movement vector~$v_i(k)$ in a distributed manner and update its own position according to 
\begin{equation}\label{eq:x_update}
x_i(k+1) = x_i(k) + v_i(k).
\end{equation}
In the proposed method, we assume that the vector~$v_i(k)$ consists of two vectors~$v_{i1}(k)$ and~$v_{i2}(k)$ as
\begin{equation}\label{eq:vector}
    v_i(k) = v_{i1}(k) + v_{i2}(k),
\end{equation}
where, as detailed below, $v_{i1}(k)$ is the vector by which the agents individually draw the desired shape, and $v_{i2}(k)$ is the vector for achieving coordination among agents.

First, we define $v_{i1}(k)$ as
\begin{equation}\label{eq:v_i1}
    v_{i1}(k) = \dot\gamma(\tau_i(k)),
\end{equation}
where $\dot\gamma$ is the derivative of function $\gamma$. We remark that, when using the vector~$v_{i1}(k)$, each agent can \emph{individually} form the desired shape. Specifically, when assuming $v_i(k)=v_{i1}(k)$, from Eqs.~\eqref{eq:x_update} and \eqref{eq:vector}, we have
\begin{equation}\label{eq:v=v1}
    x_i(k) \approx \gamma(\tau_i(k)) + y_i,
\end{equation}
where $y_i$ denotes a constant vector determined by the initial position of the agent. Consequently, we can assert that each agent can independently shape the desired formation with an offset~$y_i$. However, since the vector $y_i$ can potentially vary among agents $i=1, \dots, N$, it is not ensured that the trajectories along which each agent moves coincide. Addressing this potential discrepancy is the role of the second vector $v_{i2}(k)$. To rectify this disparity using solely the available information, agent $i$ tentatively assumes that its predecessor agent $n(i)$ is following a correct trajectory, and constructs $v_{i2}(k)$ in a manner that allows agent $i$ to align its trajectory to that of its predecessor~$n(i)$. Specifically, 
we let agent~$i$ calculate its own ``ideal position'' as     
\begin{equation}\label{eq:x_ideal}
    \tilde{x}_i(k) = x_{n(i)}\bigl(k-{(N\eta)^{-1}}\bigr).
\end{equation}
The agent~$i$ is expected to move toward this ``ideal position'' $\tilde{x}_i(k)$. For this reason, we define the movement vector~$v_{i2}$ as
\begin{equation}\label{eq:v_i2}
    v_{i2} = \alpha(\tilde{x}_i(k) - x_i(k)),
\end{equation}
where $\alpha>0$ represents a constant. Following this approach, agent 1 is anticipated to converge towards the trajectory of agent 2, while agent 2 converges towards the trajectory of agent 3, and so forth. This progression continues until agent $N$ adjusts its trajectory to converge towards the trajectory of agent 1. Consequently, it is anticipated that the trajectories of all agents will progressively align over time.

\begin{figure}[tb]
    \centering
    \includegraphics[trim={2.3cm 0 2.3cm 0},clip,width=1\textwidth]{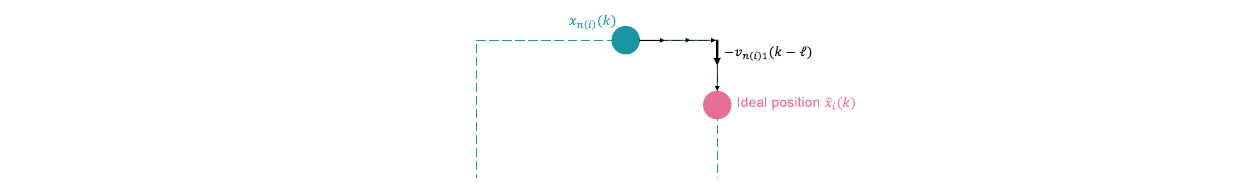}
    
    \caption{Schematic of Eq.~\eqref{eq:x_ideal}. The black arrow indicates the movement vector~$v_{n(i)1}(k-\ell)$. The agent $i$ calculates the ideal position $\tilde x_i(k)$ by repeating the computation of this vector~$v_{n(i)1}(k-\ell)$.}
    \label{fig:ideal_image}
\end{figure}

One difficulty in implementing Eq.~\eqref{eq:x_ideal} is that quantity~$x_{n(i)}$ is not directly accessible to agent~$i$. However, under the assumption that $n(i)$ is on the correct trajectory, we can approximate $\tilde{x}_i(k)$ as
\begin{equation}\label{eq:approx}
    \tilde{x}_i(k) \approx 
        x_{n(i)}(k) - \bigl(v_{n(i)1}(k-1)+v_{n(i)1}(k-2)+ \cdots  +v_{n(i)1}\bigl(k-\lfloor(N\eta)^{-1} \rfloor \bigr) \bigr),
\end{equation}
where $\flr{a}$ denotes the largest integer that does not exceed a real number $a$. Our rationale behind this approximation is illustrated in Figure~\ref{fig:ideal_image}, where the agent $i$ determines its ideal position~$\tilde{x}_i(k)$ by inverting $v_{n(i)1}(k)$, as shown by the black arrow in Figure~\ref{fig:ideal_image}. Notice that agent~$i$ can now compute the right hand side of the aforementioned approximation using only $\dot \gamma$ and $\tau_i(k)$ because a simple calculation shows
\begin{equation}
    v_{n(i)1}(k-\ell) = \dot \gamma(\tau_i(k) + N^{-1}-\ell \eta),
\end{equation}
for all $\ell \geq 1$.

\subsection{Method 2}\label{subsc:method2}

Unlike Problem 1, Problem 2 introduces the constraint that agents cannot determine their orientation as per assumption A3. Consequently, each agent must interpret the desired shape within its local coordinate system, potentially resulting in varying rotation angles $\theta_i(k)$ among agents. Therefore, achieving the desired formation hinges on aligning the rotation angles $\theta_i(k)$ of all agents. To accomplish this alignment, we propose that each agent adjusts its rotation angle to match that of its predecessor agent's coordinate system $\theta_{n(i)}(k)$. By systematically aligning their angles with those of their predecessors, it is anticipated that eventually, all agents' angles will synchronize.

The movement law for agents in Method 2 is derived from that of Method 1. Specifically, the update equation \eqref{eq:x_update} for the position $x_i(k)$, the equation \eqref{eq:vector} for the vector $v_i(k)$, and the equation \eqref{eq:v_i2} for the vector $v_{i2}(k)$ remain the same as in Method 1. However, there is a slight modification in $v_{i1}(k)$ compared to Method 1. Recall that, in Method 1, the vector $v_{i1}(k)$ was defined as in Eq. \eqref{eq:v_i1}. However, in Problem 2, as mentioned earlier, rotations of the coordinate system can occur. Therefore, the movement vector for individually forming the desired shape, $v_{i1}(k)$, needs to be of the form
\begin{equation}\label{eq:v_dassh}
    v'_{i1}(k) = 
    \begin{bmatrix}
      \cos{\theta_i(k)}&-\sin{\theta_i(k)} \\
      \sin{\theta_i(k)}&\cos{\theta_i(k)}
    \end{bmatrix}\dot\gamma(\tau_i(k))
    .
\end{equation}

In addition to the aforementioned movement law, agents undertake an action to align their orientation with the rotation angle of the predecessor agent's coordinate system $\theta_{n(i)}(k)$. However, in Problem 2, agents are assumed to lack direct access to the angle of the predecessor agent $\theta_{n(i)}(k)$. Instead, leveraging the available information—namely, the relative position of the predecessor agent between two steps—we endeavor to estimate the angle of the predecessor agent $\theta_{n(i)}(k)$.
Specifically, the agent~$i$ first constructs an estimation of the movement vector of the predecessor agent as
\begin{equation}\label{eq:vector_ni}
    v_{n(i)}(k) = x_{n(i)}(k) - x_{n(i)}(k-1).
\end{equation}
The computation of this vector is possible for agent~$i$ because, according to the problem formulation, agent~$i$ can obtain the relative position of the predecessor agent $n(i)$ (i.e. $x_{n(i)}(k) - x_i(k)$ and $x_{n(i)}(k-1) - x_i(k)$). 
Next, we observe that the following equation holds true from Eq.~\eqref{eq:v_dassh}:
\begin{equation}\label{eq:v_dassh_tau}
    v'_{n(i)1}(k) = 
    \begin{bmatrix}
      \cos{\theta_{n(i)}(k)} & -\sin{\theta_{n(i)}(k)} \\
      \sin{\theta_{n(i)}(k)} &\cos{\theta_{n(i)}(k)}
    \end{bmatrix}\dot\gamma(\tau_{n(i)}(k))
    .
\end{equation}
Notice that the agent~$i$ can calculate the value of $\tau_{n(i)}(k)$ of the predecessor agent because Eqs.~\eqref{eq:tau_update} and~\eqref{eq:distance1/N} indicate
\begin{equation}\label{eq:tau_ni}
\tau_{n(i)}(k) =
\begin{cases}
\displaystyle 
\tau_i(k) + 1/N
, &\mbox{if $\tau_i(k) + 1/N < 1$}, 
\\
\displaystyle 
\tau_i(k) + 1/N - 1
, &\mbox{otherwise},
\end{cases}
\end{equation}
at every step $k$.
Therefore, under the approximation $v_{n(i)}(k) \approx v_{n(i)1}(k)$, from \eqref{eq:vector_ni} we obtain 
\begin{equation}\label{eq:theta_ni}
    x_{n(i)}(k) - x_{n(i)}(k-1) \approx 
    \begin{bmatrix}
        \cos{\theta_{n(i)}(k)} & -\sin{\theta_{n(i)}(k)} \\
        \sin{\theta_{n(i)}(k)} & \cos{\theta_{n(i)}(k)}
    \end{bmatrix}\dot\gamma(\tau_{n(i)}(k)), 
\end{equation}
which allows the agent $i$ to compute an estimation~$\tilde\theta_{n(i)}$ of $\theta_{n(i)}$, as both $x_{n(i)}(k) - x_{n(i)}(k-1)$ and $\dot\gamma(\tau_{n(i)}(k))$ are computable for the agent~$i$.
The impact of this approximation will be discussed in Section~\ref{sc:sim}.

In this study, agents are allowed to adjust their rotation angles randomly at each time step. In this paper, we propose various methods for dynamically adjusting the probability $\beta$, as summarized in Table~\ref{tab:beta_method_name}. Firstly, the Constant method fixes $\beta$ as a constant value. Next, the Time method varies $\beta$ as the number of steps progresses. In the Time-decrease method, $\beta$ gradually decreases, whereas in the Time-increase method, $\beta$ gradually increases. The third method, Achievement, adjusts $\beta$ based on the formation achievement rate, as will be elaborated later. In the Achievement-decrease method, a higher formation achievement rate leads to a smaller $\beta$, while in the Achievement-increase method, a higher formation achievement rate leads to a larger $\beta$.
\begin{table}[tb]
    \centering
    \caption{Summary of methods of adjusting probability~$\beta$.}
    \label{tab:beta_method_name}
    \begin{tabular}{cc}
        \hline
        Name & How to determine the value $\beta$ \\
        \hline
        Constant & Constant value \\
        \hline
        \begin{tabular}{c}
            Time--decrease \\
            Time--increase
        \end{tabular}
        & According to steps \\
        \hline
        \begin{tabular}{c}
            Achievement--decrease \\
            Achievement--increase
        \end{tabular} &  According to formation achievement \\
        \hline
    \end{tabular}
\end{table}

We delineate the calculation method for the formation achievement rate within the aforementioned Achievement method. Notably, agents are tasked with computing the formation achievement rate utilizing only the limited information outlined in Section~\ref{sc:statement}. Accordingly, in this paper, the magnitude of the difference in orientation angles between each agent and its predecessor serves as the indicator for the degree of formation achievement. Thus, the achievement rate of agent $i$ at step $k$ is defined as:
\begin{equation}\label{eq:achievement}
A_i(k) = \abs{\theta_i(k) - \tilde\theta_{n(i)}(k)}.
\end{equation}
From Eq.~\eqref{eq:achievement}, the achievement rate $A_i(k)$ decreases as the difference in orientation with the leading agent's coordinate system decreases. Therefore, the probability $\beta$ is determined to be proportional to $A_i(k)$ in the Achievement--decrease method and inversely proportional in the Achievement--increase method. In other words, for the Achievement--decrease method, we set
\begin{equation}\label{eq:Ad}
    \beta = c_d A_i(k),
\end{equation}
while for the Achievement--increase method, we set 
\begin{equation}\label{eq:Ai}
    \beta = \frac{c_i}{A_i(k)},
\end{equation}
where $c_d$ and $c_i$ are positive constants.
\section{Numerical Simulations}\label{sc:sim}

In this section, we assess the effectiveness of the proposed method. Initially, in Section~\ref{subsc:sim_settings}, we outline the simulation settings employed in this study. Subsequently, in Section~\ref{subsc:qualitative}, we present the trajectories of agents to demonstrate the efficacy of the proposed methods. Following this qualitative analysis, in Section~\ref{subsc:quantitative}, we conduct a quantitative evaluation of the methods' effectiveness. Finally, in Sections~\ref{subsc:varying_N} and \ref{subsc:other_shape}, we alter certain settings and explore the versatility of the methods.

\subsection{Simulation Settings}
\label{subsc:sim_settings}

In this section, we delineate the simulation settings utilized in this study. As outlined in Section~\ref{sc:method}, we assumed that agents are positioned equidistantly on a circle at the initial step $k=0$. The circle's center and radius are established at the origin and $15$, respectively. The number of agents $N$ varied, with simulations employing 3 agents in Sections~\ref{subsc:qualitative}, \ref{subsc:quantitative}, and \ref{subsc:other_shape}, and 5 to 30 agents in Section~\ref{subsc:varying_N}. The constant $\eta$ in Eq.~\eqref{eq:tau_update} was fixed at $0.01$, and $\alpha$ in Eq.~\eqref{eq:v_i2} was set to $0.01$. Moreover, the maximum simulation step was capped at $1000$. Furthermore, in Method 2's Constant method, $\beta$ is initialized to 0.1. For the Time-decrease and Time-increase methods of Method 2, $\beta$ commences at 0.1 in the initial step $k=0$, and then decreases and increases by 0.01 every 100 steps, respectively. In the Achievement-decrease method, the coefficient $c_d$ in Eq.~\eqref{eq:Ad} is adjusted to $(10\pi)^{-1}$, while in the Achievement-increase method, the coefficient $c_i$ in Eq.~\eqref{eq:Ai} is set to $0.01\pi$. 
Additionally, the probability~$\beta$ in the Achievement-increase method is initialized at the value of~$0.1$, while in the Achievement-decrease method, it begins at $0.01$. 

\subsection{Qualitative Evaluation}\label{subsc:qualitative}

\begin{figure}[tb]
\centering
  \includegraphics[trim={2.3cm 0 2.3cm 0},clip,width=1\textwidth]{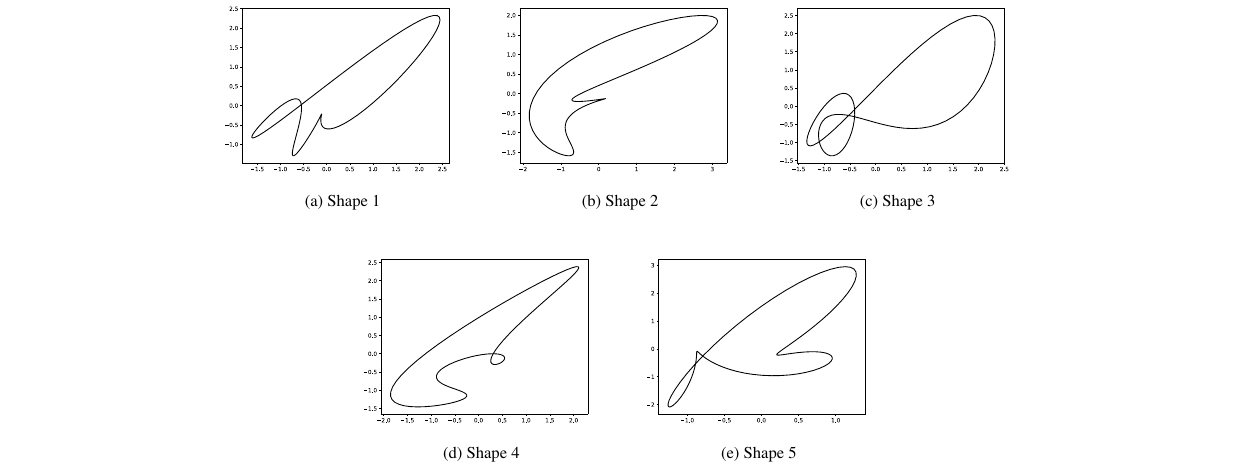}
  \caption{Five desired shapes generated from Fourier series~\eqref{eq:fourier}.}
  \label{fig:desired}
\end{figure}

In this section, we conduct a qualitative assessment of the proposed methods' effectiveness. Initially, we outline the process for determining the desired shapes. The desired shape $\gamma$ is represented in the form of a Fourier series. To ensure that $\gamma$ forms a closed curve, it is defined as
\begin{equation}\label{eq:fourier}
    \gamma(t) = 
    \begin{bmatrix}
        \displaystyle 
        c_{x0} + c_{x1} \sin{(2\pi t)} + c_{x2} \cos{(2\pi t)} + c_{x3}\sin{(4\pi t)} + c_{x4}\cos{(4\pi t)}+ \cdots
        \\
        \displaystyle 
        c_{y0} + c_{y1} \sin{(2\pi t)} + c_{y2} \cos{(2\pi t)} + c_{y3}\sin{(4\pi t)} + c_{y4}\cos{(4\pi t)}+ \cdots
    \end{bmatrix}
    ,
\end{equation}
where $c_{x1}, c_{x2}, \dotsc$ and $c_{y1}, c_{y2}, \dotsc$ denote constants. For simplicity, we assume $c_{x0} = c_{y0} = 0$. We generated five desired shapes (refer to Figure~\ref{fig:desired}) by randomly generating the coefficients, as shown in Table~\ref{tab:coefficient}. It's worth noting that, to avoid overly complex shapes, we set the coefficients $c_{x1}$ and $c_{y2}$ to $1$.

\begin{table}[tb]
    \centering
    \caption{Combination of the coefficients in the Fourier series~\eqref{eq:fourier} for the figures shown in Figure~\ref{fig:desired}.}
    \label{tab:coefficient}
    \subfloat[$\gamma_x(t)$]{
    \centering
    \begin{tabular}{c|ccccc}
         & 1 & 2 & 3 & 4 & 5 \\
        \hline
        $c_{x0}$ & 0 & 0 & 0 & 0 & 0 \\
        $c_{x1}$ & 1 & 1 & 1 & 1 & 1 \\
        $c_{x2}$ & 0.76 & 0.85 & 0.95 & 0.54 & 0.10 \\
        $c_{x3}$ & 0.42 & 0.76 & 0.06 & 0.37 & 0.40 \\
        $c_{x4}$ & 0.26 & 0.26 & 0.08 & 0.60 & 0.15 \\
        $c_{x5}$ & 0.51 & 0.50 & 0.84 & 0.63 & 0.07 \\
        $c_{x6}$ & 0.40 & 0.45 & 0.74 & 0.07 & 0.40
    \end{tabular}
    \label{subtab:gamma_x}
    }
    \subfloat[$\gamma_y(t)$]{
    \begin{tabular}{c|ccccc}
         & 1 & 2 & 3 & 4 & 5 \\
        \hline
        $c_{y0}$ & 0 & 0 & 0 & 0 & 0 \\
        $c_{y1}$ & 0.30 & 0.79 & 0.31 & 0.84 & 0.80 \\
        $c_{y2}$ & 1 & 1 & 1 & 1 & 1 \\
        $c_{y3}$ & 0.58 & 0.03 & 0.61 & 0.23 & 0.22 \\
        $c_{y4}$ & 0.45 & 0.94 & 0.61 & 0.26 & 0.77 \\
        $c_{y5}$ & 0.50 & 0.43 & 0.16 & 0.47 & 0.28 \\
        $c_{y6}$ & 0.91 & 0.84 & 0.58 & 0.99 & 0.54 \\
    \end{tabular}
    \label{subtab:gamma_y}
    }
\end{table}

\begin{figure}[tb]
    \centering
    \includegraphics[trim={2.3cm 0 2.3cm 0},clip,width=1\textwidth]{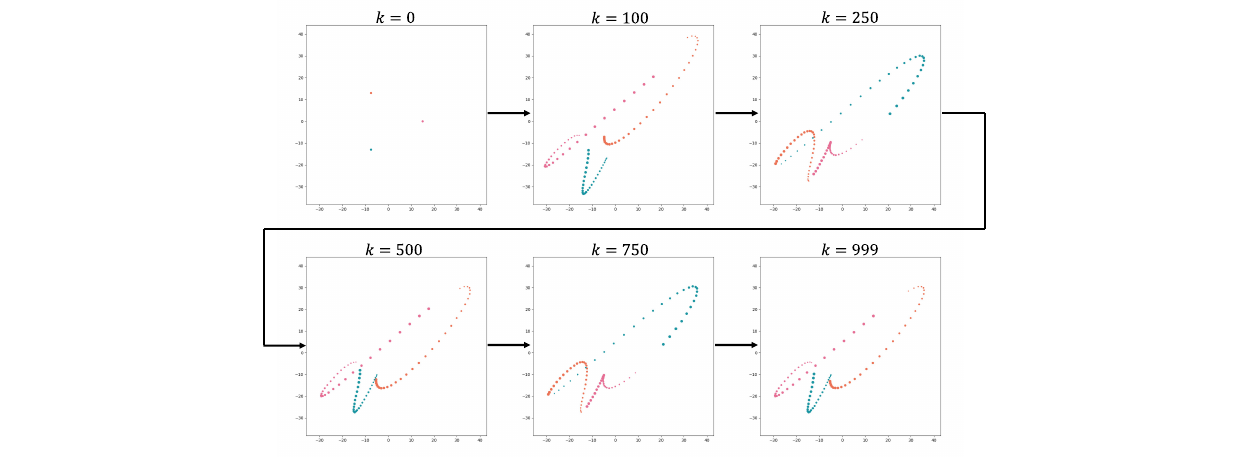}
    \caption{The agents' trajectories for the desired shape~1 when using Method~1.}
    \label{fig:trajectory_method1-0}
\end{figure}

Initially, we present snapshots from the simulation results of Method 1 in Figure~\ref{fig:trajectory_method1-0}. In this figure, the trajectories of each agent for the past 30 steps at $k=0, 100, 250, 500, 750, 999$ are displayed from top left to top right, and then from bottom left to bottom right. It is evident from this figure that the agents progressively trace the desired shape. Moreover, this trend was consistently observed across the other four desired shapes.

\begin{figure}[tb]
    \centering
    \includegraphics[trim={2.3cm 0 2.3cm 0},clip,width=1\textwidth]{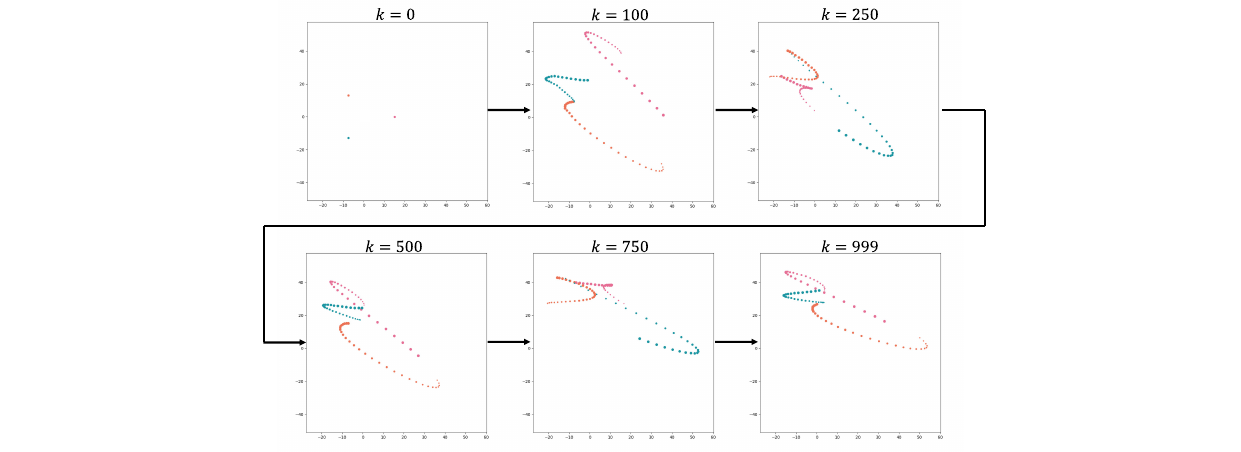}
    \caption{The agents' trajectory for each desired shape when using Method~2, Constant.}
    \label{fig:trajectory_C-0}
\end{figure}

\begin{figure}[tb]
    \centering
    \includegraphics[trim={2.3cm 0 2.3cm 0},clip,width=1\textwidth]{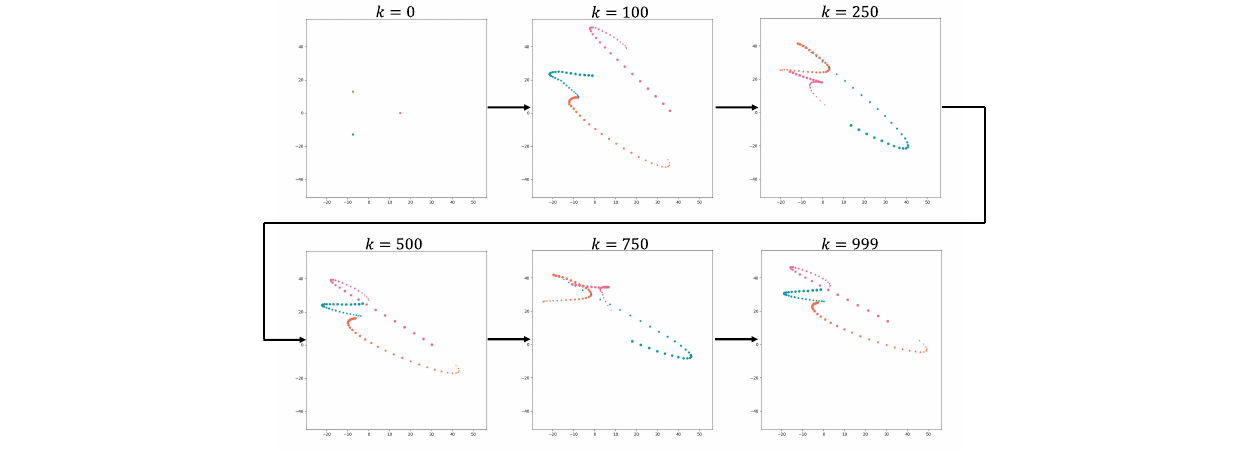}
    \caption{The agents' trajectory for each desired shape when using Method~2, Time--decrease.}
    \label{fig:trajectory_Td-0}
\end{figure}

\begin{figure}[tb]
    \centering
    \includegraphics[trim={2.3cm 0 2.3cm 0},clip,width=1\textwidth]{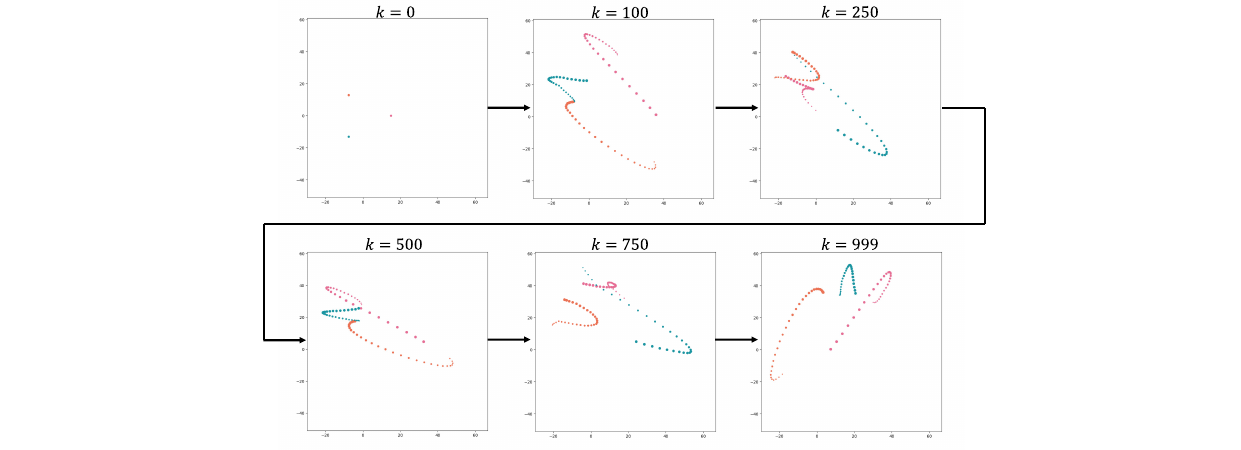}
    \caption{The agents' trajectory for each desired shape when using Method~2, Time--increase.}
    \label{fig:trajectory_Ti-0}
\end{figure}

\begin{figure}[tb]
    \centering
    \includegraphics[trim={2.3cm 0 2.3cm 0},clip,width=1\textwidth]{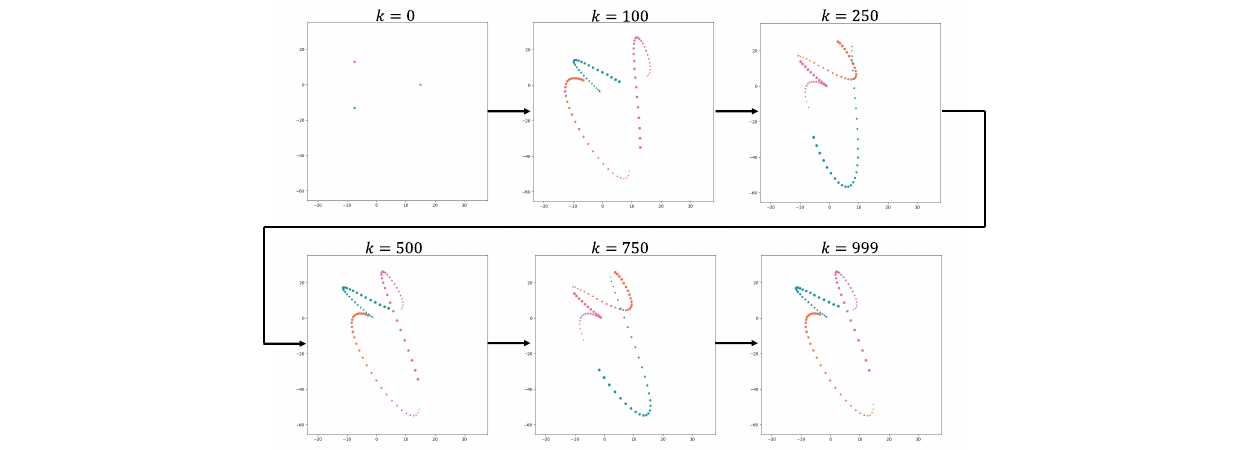}
    \caption{The agents' trajectory for each desired shape when using Method~2, Achievement--decrease.}
    \label{fig:trajectory_Ad-0}
\end{figure}

\begin{figure}[tb]
    \centering
    \includegraphics[trim={2.3cm 0 2.3cm 0},clip,width=1\textwidth]{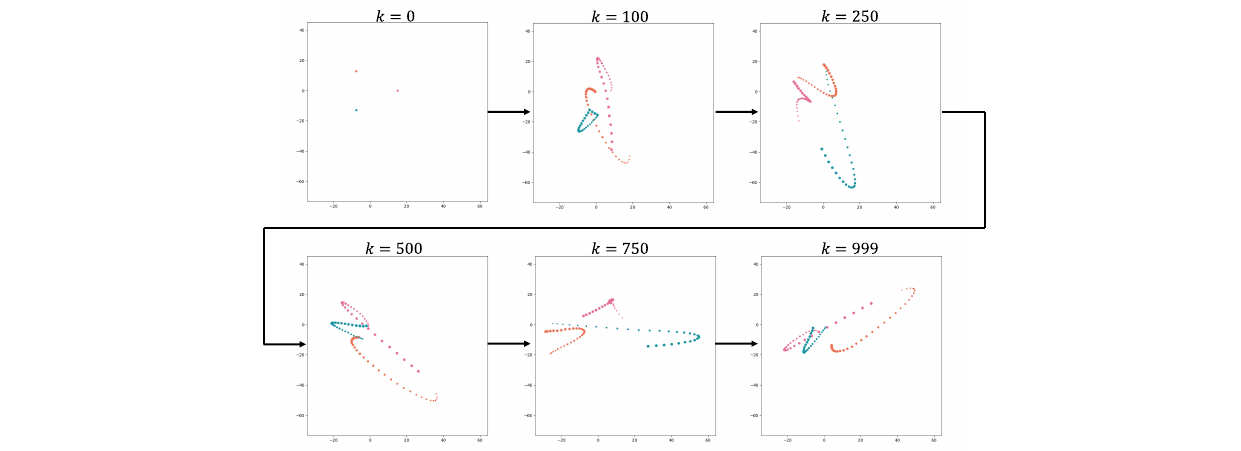}
    \caption{The agents' trajectory for each desired shape when using Method~2, Achievement--increase.}
    \label{fig:trajectory_Ai-0}
\end{figure}

Next, we present the results of Method 2. Figures~\ref{fig:trajectory_C-0} to \ref{fig:trajectory_Ai-0} showcase the trajectories of the agents when employing different methods for setting the probability~$\beta$ (refer to Table~\ref{tab:beta_method_name}). In these figures, Shape 1 is chosen as the desired shape. It is evident from these visualizations that distinct outcomes were achieved for each method. Notably, the Achievement-decrease method appears to achieve the formation most effectively. Specifically, in this method, the formation is nearly attained by $k=250$, and it remains stable thereafter. Conversely, in the other four methods, although the formation initially seems to be achieved, significant breakdowns occur afterward (e.g., Figure~\ref{fig:trajectory_C-0}, step $k=750$).
The observed phenomenon, peculiar to Method 2, is attributed to the approximation error of the vector $v_{n(i)1}(k)$ in Eq.~\eqref{eq:vector_ni}. Likewise, the error in estimating the orientation of the predecessor agent's coordinate system also stems from this approximation error. Consequently, formations that appear to have been achieved may experience significant collapses subsequently. Further investigation into this formation collapse will be undertaken in Section~\ref{subsc:quantitative}.


\subsection{Quantitative Evaluation}\label{subsc:quantitative}

In this section, we undertake a quantitative assessment of the accuracy of our formation control approach. One aspect of this evaluation involves demonstrating the fulfillment of Eq.~\eqref{eq:position_lim}. However, it's important to acknowledge that obtaining the shape $\tilde\gamma$ referenced in Eq.~\eqref{eq:position_lim} requires enlarging and translating the desired shape $\gamma$ for Problem 1, and additionally rotating it for Problem 2. As such, acquiring $\tilde\gamma$ is not a straightforward task, suggesting that directly evaluating Eq.~\eqref{eq:position_lim} may not be entirely feasible. Therefore, instead, we propose an alternative evaluation metric in Section~\ref{subsc:minimize_Frechet}, utilizing it to quantitatively assess the effectiveness of our proposed method.

\begin{figure}[tb]
    \centering
    \includegraphics[trim={2.3cm 0 2.3cm 0},clip,width=1\textwidth]{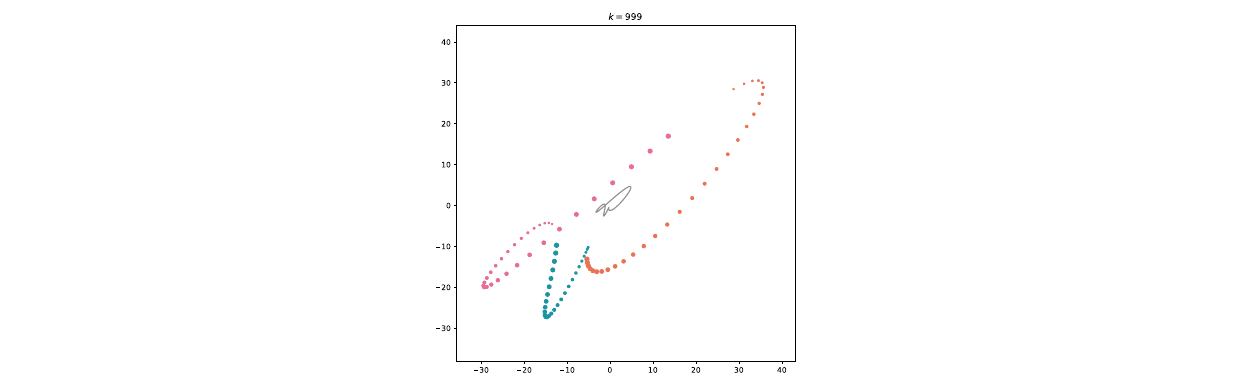}
    \caption{An example of the trajectories of agents. The gray line represents the desired shape $\gamma$, and the dots indicate the trajectories traced by the three agents. Although the agents seem to achieve the formation, their trajectories do not align with those of the desired shape $\gamma$.}
    \label{fig:diff_of_shapes}
\end{figure}

\subsubsection{Evaluation metric}\label{subsc:minimize_Frechet}

The proposed evaluation metric is based on the Fréchet distance~\cite{eiter1994computing}. Our aim is to calculate the Fréchet distance between the trajectories of the agents and the desired shape. However, directly computing the distance with the desired shape $\gamma$ may not yield an appropriate evaluation. For instance, as illustrated in Figure~\ref{fig:diff_of_shapes}, while the trajectories of the agents achieve the formation, their shapes may differ from the original shape $\gamma$ due to enlargement and translation. Consequently, the Fréchet distance with the original shape $\gamma$ might remain large. To mitigate this issue, we consider deforming the desired shape $\gamma$ to minimize the Fréchet distance and evaluate the distance at that instance.
Specifically, we adopt the metric defined as follows to evaluate the performance of the proposed methods: 
\begin{equation}\label{eq:ev_func}
    d(s) = \min_{e, x_t, y_t\in\mathbb{R}, \theta\in [0,2\pi]}{\frac{1}{N}\sum_{i=0}^{N-1}d_F(\gamma_i(s), D_i(e,\theta,x_t,y_t))},
\end{equation}
where $d_F(a,b)$ denotes the Fréchet distance between shapes $a$ and $b$, $\gamma_i(s)$ represents the trajectory of agent $i$ during a period $s$ (defined as the number of steps required for an agent to complete one cycle of the desired shape, i.e., $\eta^{-1}$), and $D(e,\theta,x_t,y_t)$ indicates the trajectory obtained under scaling factor $e$, rotation angle $\theta$, and parallel translation to the point $(x_t, y_t)$ for the desired shape. While obtaining the exact value of the metric $d$ may not be straightforward, the use of a genetic algorithm enabled us to approximately compute the metric with a certain level of efficiency.

\subsubsection{Simulation Results}

\begin{figure}[tb]
    \centering
    \includegraphics[trim={2.3cm 0 2.3cm 0},clip,width=1\textwidth]{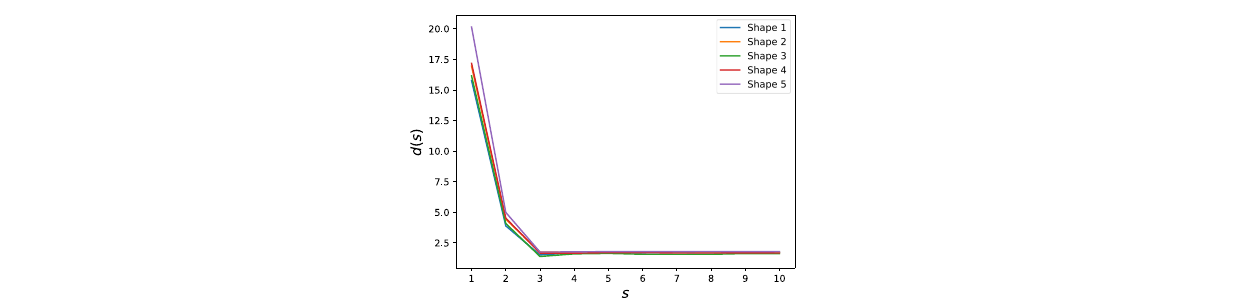}
    \caption{Time series of the metric~$d$ when using Method~1. The horizontal axis represents the period.}
    \label{fig:frechet_method1}
\end{figure}

In this section, we conduct a quantitative evaluation based on the metric introduced in Section~\ref{subsc:minimize_Frechet}.
Firstly, we present the results obtained using Method~1 in Figure~\ref{fig:frechet_method1}. Here, the horizontal axis represents the period, while the vertical axis indicates the metric~$d$.
From this figure, it is evident that for all five shapes, the average Fréchet distance of the agents decreases over time, eventually reaching and maintaining a small value as the period progresses. This consistent trend suggests the successful achievement of the formation.

\begin{figure}[tb]
    \centering
    \includegraphics[trim={2.3cm 0 2.3cm 0},clip,width=1\textwidth]{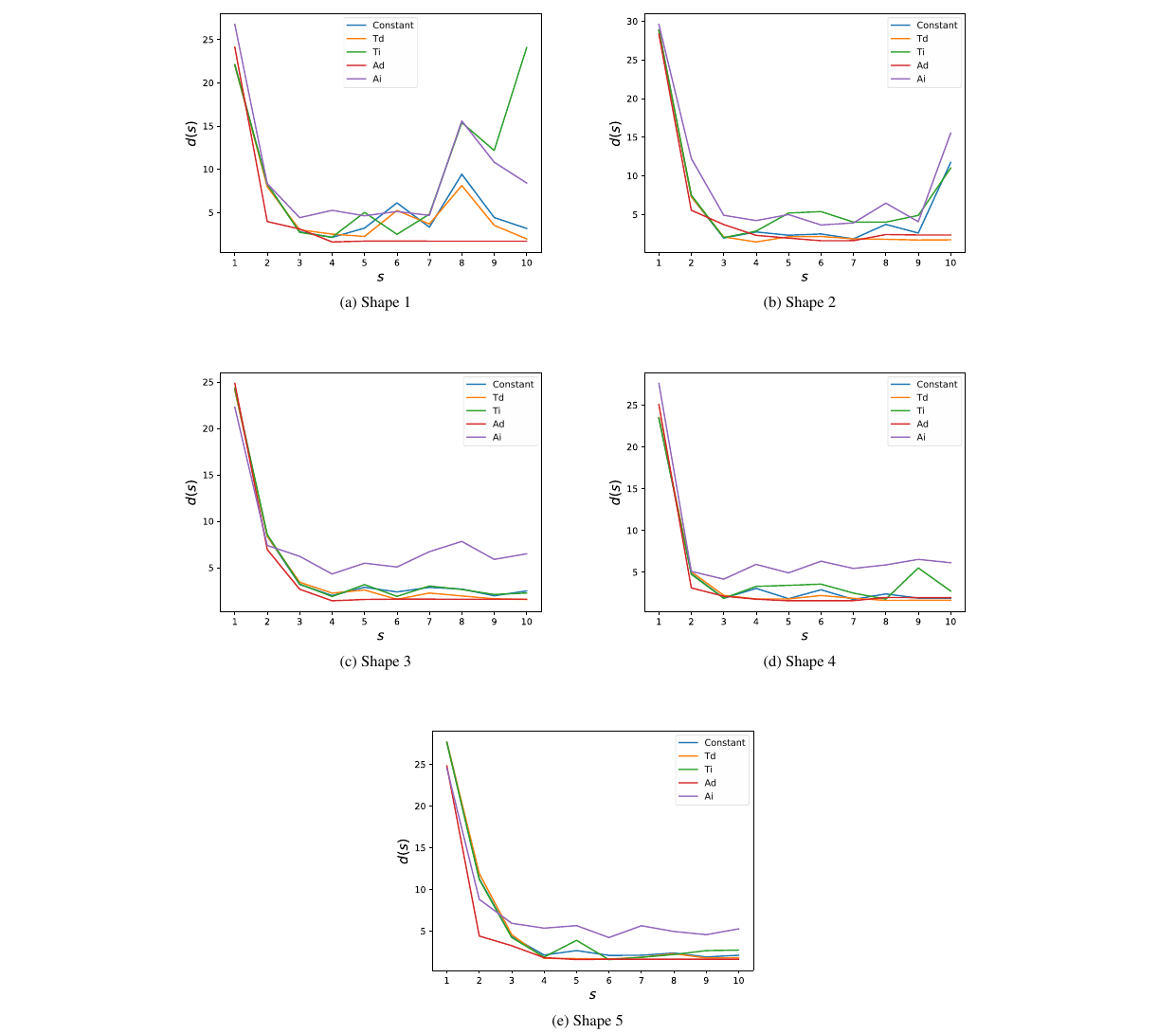}
    \caption{Time series of the metric~$d$ when using Method~1. The horizontal axis represents the period. Labels Td, Ti, Ad, Ai denote the Time--decrease, Time--increase, Achievement--decrease and Achievement--increase method, respectively.}
    \label{fig:frechet_method2}
\end{figure}

Next, we present the results obtained using Method~2. In Figure~\ref{fig:frechet_method2}, we depict the time series of the metric when employing different $\beta$ methods (See Table~\ref{tab:beta_method_name}).
Across all methods and shapes, it is evident that the metric consistently decreases compared to the initial period $s=1$. However, the choice of the probability~$\beta$ significantly influences the degree of decrease in the metric. Notably, the Achievement--decrease method (Fig.~\ref{fig:trajectory_Ad-0})consistently maintains a low metric for all five shapes, unlike other methods that exhibit sudden increases in the metric. This phenomenon is most pronounced in Shape~1, where all methods, except for the Achievement--decrease, experience a notable increase in the metric. 
This behavior is attributed to the approximation error in the vector $v_{n(i)1}(k)$ of the predecessor agent, as discussed in Section~\ref{subsc:qualitative}. Specifically, errors stemming from the approximation of~$v_{n(i)1}(k)$ can lead to inaccuracies in estimating the rotation angle of the predecessor agent's coordinate system. To mitigate this issue, it is advisable to reduce the probability~$\beta$ of executing the rotation of the coordinate system when the formation is achieved. This hypothesis finds support in the Achievement--decrease method, where the probability~$\beta$ decreases upon formation achievement, resulting in no formation collapse.

To further validate this hypothesis, we examine the relationship between the values of~$\beta$ and the metric for each method. Focusing on Shape~1, where formation collapse was most evident, Table~\ref{tab:beta} shows the time series of the magnitude of~$\beta$ for each method.
Upon inspection, it is notable that the Achievement--decrease method, exhibiting the smallest metric, demonstrates a decreasing trend in the magnitude of the probability~$\beta$ as the cycles progress. Conversely, the Achievement--increase method maintains a constant probability~$\beta$ value of 1 throughout. However, no significant difference in formation achievement is observed compared to the other three methods, where a phenomenon of metric decrease followed by an increase at a certain cycle is consistent.
Based on these observations, and in alignment with the hypothesis, it is inferred that reducing the probability~$\beta$ after formation achievement is beneficial. Additionally, maintaining the probability below 0.1 at all times is deemed favorable.
\begin{table}[tb]
    \centering
    \caption{Time series of the probability~$\beta$ averaged over each periods. Labels C, Td, Ti, Ad, and Ai denote the Constant, Time--decrease, Time--increase, Achievement--decrease, and Achievement--increase methods, respectively.%
    }
    \label{tab:beta}
    \begin{tabular}{ccccccccccc}
         & 1 & 2 & 3 & 4 & 5 & 6 & 7 & 8 & 9 & 10 \\
        \hline
        C & 0.1 & 0.1 & 0.1 & 0.1 & 0.1 & 0.1 & 0.1 & 0.1 & 0.1 & 0.1 \\
        Td & 0.1 & 0.09 & 0.08 & 0.07 & 0.06 & 0.05 & 0.04 & 0.03 & 0.02 & 0.01 \\
        Ti & 0.1 & 0.11 & 0.12 & 0.13 & 0.14 & 0.15 & 0.16 & 0.17 & 0.18 & 0.19  \\
        Ad & 0.02 & 0.004 & 0.002 & $10^{-3}$ & $10^{-3}$ & $10^{-3}$ & $10^{-3}$ & $10^{-3}$ & $10^{-3}$ & $10^{-3}$ \\
        Ai & 1 & 1 & 1 & 1 & 1 & 1 & 1 & 1 & 1 & 1 \\
    \end{tabular}
\end{table}

\begin{figure}[tb]
    \includegraphics[trim={2.3cm 0 2.3cm 0},clip,width=1\textwidth]{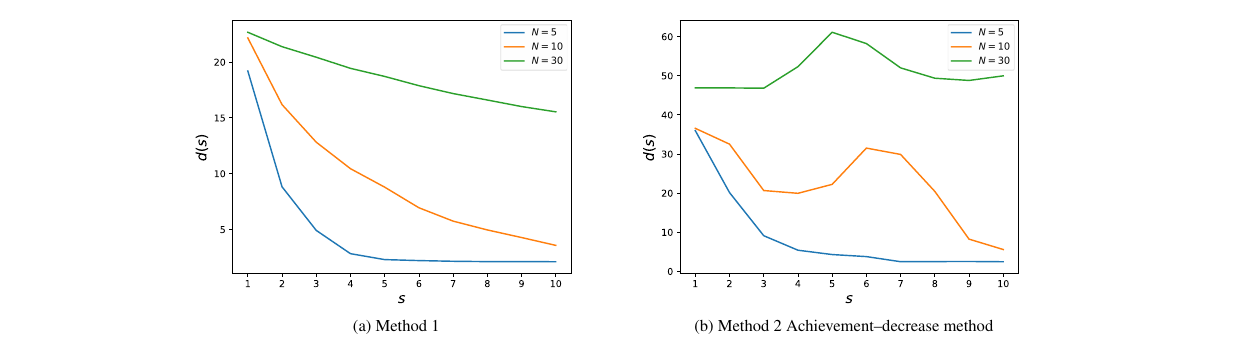}
    \caption{Time series of the metric~$d$ for the case of $5$, $10$, and $30$ agents. The horizontal axis represents the period.}
    \label{fig:frechet_N}
\vspace{5mm}
    \includegraphics[trim={2.3cm 0 2.3cm 0},clip,width=1\textwidth]{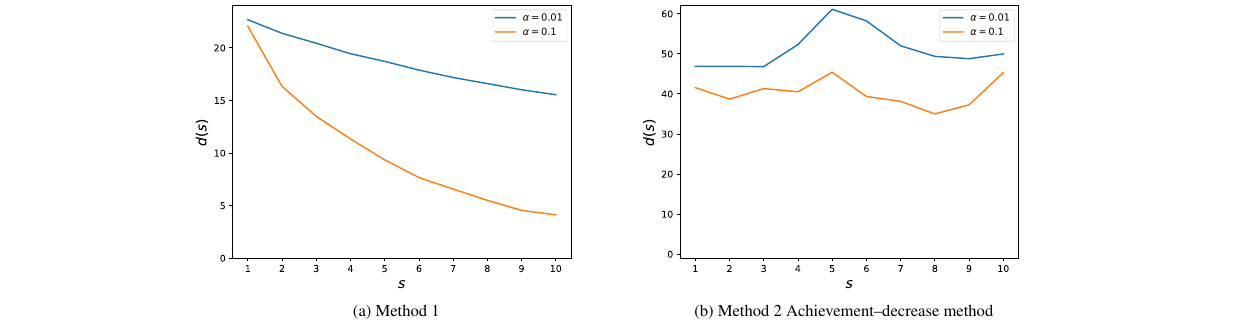}
    \caption{Time series of the metric~$d$ for the case of $30$ agents. The horizontal axis represents the period.}
    \label{fig:frechet_N-alpha}
\end{figure}

\subsection{Varying the Number of Agents}\label{subsc:varying_N}

We conducted simulations with varying agent counts ($N$) to assess the effectiveness of the proposed method concerning the number of agents.
For Method~2, we choose to use the Achievement--decrease method as it showed superior performance in the numerical simulations presented so far.
Figure~\ref{fig:frechet_N} illustrates the metric for $N=5, 10,$ and $30$. It becomes apparent that as the number of agents increases, achieving the formation becomes more challenging. This trend was observed in both Method 1 and Method 2, indicating a shared cause related to the movement vectors. In scenarios with numerous agents, predecessor positions are relatively closer, reducing the magnitude of the term~$v_{i2}$. Consequently, an imbalance in the vectors $v_{i1}(k)$ and $v_{i2}(k)$ in Eq.~\eqref{eq:vector} emerges, posing difficulties in achieving formation.

To test this hypothesis, simulations were conducted by varying the magnitude of the coefficient $\alpha$ for the case of~$N=30$, and the impact on formation achievement was investigated.
In Figure~\ref{fig:frechet_N-alpha}, we show the results of the simulation when changing the value $\alpha$.

In both methods, it can be observed that the metric is smaller when $\alpha=0.1$. This observation supports the aforementioned hypothesis. However, particularly for Method~2, the metric is not sufficiently small. Developing a method that can achieve formation adequately even with a large number of agents is a potential future work.

\begin{figure}[tb]
\centering
    \includegraphics[trim={2.3cm 0 2.3cm 0},clip,width=1\textwidth]{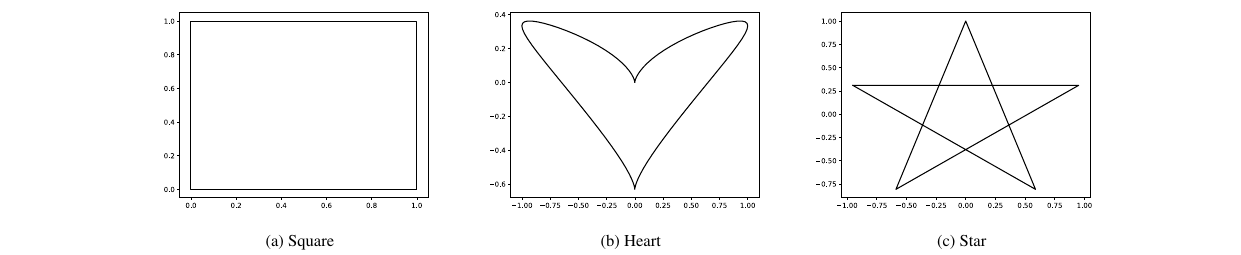}
    \caption{Three additional desired shapes.}
    \label{fig:desired_hatten}
\vspace{5mm}
\centering
    \includegraphics[trim={2.3cm 0 2.3cm 0},clip,width=1\textwidth]{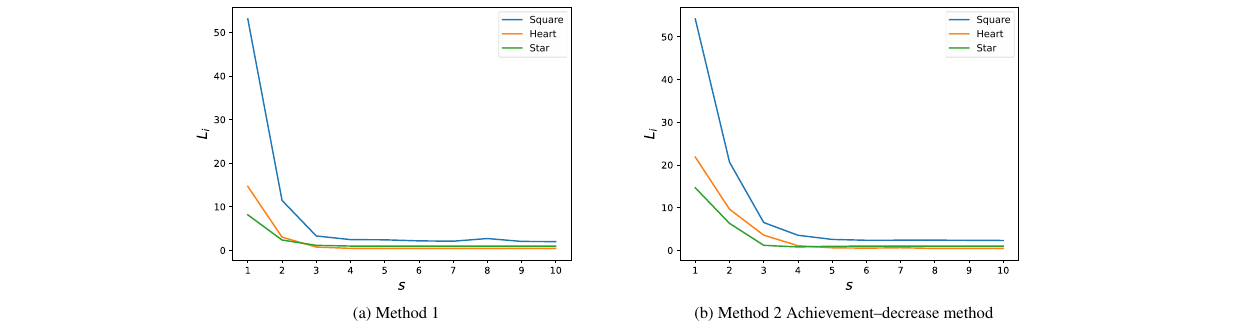}
    \caption{Time series of the metric~$d$ for Square, Heart, and Star, respectively. The horizontal axis represents the period.}
    \label{fig:frechet_hatten}
\end{figure}

\subsection{Other Shapes}\label{subsc:other_shape}

Let us further investigate the effectiveness of the proposed method using more complex shapes with multiple angles and intersections. The shapes used in this section are a square (Figure~\ref{fig:desired_hatten}a), a heart shape (Figure~\ref{fig:desired_hatten}b) and a star shape (Figure~\ref{fig:desired_hatten}c).
Here, a square is a shape passing through the points $[0,0]^\top$, $[1,0]^\top$, $[1,1]^\top$, $[0,1]^\top$ in order, a star is a shape passing through the points $[\cos{(9\pi/10)}, \sin{(9\pi/10)}]^\top$, $[\cos{(\pi/10)}, \sin{(\pi/10)}]^\top$, $[\cos{(13\pi/10)}, \sin{(13\pi/10)}]^\top$, $[\cos{(\pi/2)}, \sin{(\pi/2)}]^\top$, $[\cos{(17\pi/10)}, \sin{(17\pi/10)}]^\top$ in order, and a heart is a shape given as
\begin{equation}
    \gamma(t)= 
    \begin{bmatrix}
        \sin^3{2\pi t} \\
        (\cos{2\pi t} - \cos{4\pi t}) / (3-(\cos{6\pi t}/6))
    \end{bmatrix}.
\end{equation}

In Method 2, the approach for determining $\beta$ is set to be the Achievement--decrease method.
In Figure~\ref{fig:frechet_hatten}, the results of the formation in each method are presented. From these two figures, it is suggested that in both methods, the formation of three additional shapes was successfully achieved. Thus, it can be inferred that formations of more complex shapes, including angles and intersections, can also be accomplished.

We remark that the two proposed methods do not take into consideration collisions between agents. Therefore, especially for shapes with intersections such as the Star shape, there is a need for the development of a method that can achieve formations while avoiding collisions. Even in shapes without intersections, collisions are expected during the movement toward formation achievement. Hence, it is crucial to develop a method that can effectively avoid collisions. This remains a future research topic.
\section{Conclusion}\label{sc:conclusion}

In this study, we presented movement laws for agents performing cyclic pursuit to autonomously form various shapes, not limited to circles, ellipses, and figure-eights. We addressed two distinct problem settings to achieve this goal.
In the first setting, we devised a system where agents contribute to shape formation through two vectors: $v_{i1}$ representing individual formation and $v_{i2}$ guiding them towards their predecessor agents. This approach aimed at collectively forming the desired shape. In the second setting, considering scenarios where agents lack shared coordinate systems, we introduced additional rotations to align their coordinate systems with those of their predecessors, alongside the previously mentioned vectors.

Simulation results demonstrated the effectiveness of our proposed method. Agents progressively moved to trace desired shapes, with a gradual decrease observed in the discrepancy between desired shapes and agent trajectories, measured using the Fréchet distance. Notably, Method 2, particularly the Achievement-decrease approach, exhibited superior performance by reducing the probability of coordinate system rotations once the formation was achieved. Moreover, the method successfully achieved formations even for complex shapes with multiple angles and intersections.

For future research, it's crucial to explore the generality of our method by varying parameters beyond the number of agents. This includes investigating formation achievement rates with modifications in parameters other than the agent count and examining the method's robustness against noise. Additionally, developing strategies for formations involving a large number of agents and proposing collision avoidance behaviors are essential future challenges. Furthermore, while this study focused on providing sufficient information to agents for shape formation, future efforts should aim to determine the minimum information required for arbitrary shape formation.

\section*{Acknowledgement}

This work was partially supported by JSPS KAKENHI Grant Number JP21H01352.





\end{document}